\documentclass[pra,preprint,showpacs,showkeys]{revtex4}
\usepackage{graphicx}
\usepackage{amsmath}
\usepackage{amssymb}
\usepackage{dsfont}

\newtheorem{theorem}{Theorem}

\newcommand{\ket}[1]{\left\vert#1\right\rangle} 
\newcommand{\bra}[1]{\left\langle#1\right\vert} 
\newcommand{\tr}{\mathrm{tr}}

\begin{document}

\title{Heat Transfer Operators Associated with Quantum Operations}
\author{\c{C}. Aksak}
\author{S. Turgut}
\affiliation{
Department of Physics, Middle East Technical University,\\
TR-06800, Ankara, Turkey}

\begin{abstract}
Any quantum operation applied on a physical system is performed as
a unitary transformation on a larger extended system. If the
extension used is a heat bath in thermal equilibrium, the
concomitant change in the state of the bath necessarily implies a
heat exchange with it. The dependence of the average heat
transferred to the bath on the initial state of the system can
then be found from the expectation value of a hermitian operator,
which is named as the heat transfer operator (HTO). The purpose of
this article is the investigation of the relation between the HTOs
and the associated quantum operations. Since, any given quantum
operation on a system can be realized by different baths and
unitaries, many different HTOs are possible for each quantum
operation. On the other hand, there are also strong restrictions
on the HTOs which arise from the unitarity of the transformations.
The most important of these is the Landauer erasure principle.
This article is concerned with the question of finding a complete
set of restrictions on the HTOs that are associated with a given
quantum operation. An answer to this question has been found only
for a subset of quantum operations. For erasure operations, these
characterizations are equivalent to the generalized Landauer
erasure principle. For the case of generic quantum operations
however, it appears that the HTOs obey further restrictions which
cannot be obtained from the entropic restrictions of the
generalized Landauer erasure principle.
\end{abstract}



\maketitle

\section{Introduction}

A fundamental result in the thermodynamics of computation is
Landauer's erasure principle (LEP), which places a lower bound on
the average heat dumped to a bath when the information in a memory
device is erased. The principle holds for processes that satisfy
two essential features: The process is carried out in the same way
independent of the information stored in the device and it
restores the device to a known \emph{standard state} at the end.
These processes are usually known as the \emph{Landauer erasure}
processes. Landauer has shown that if the device holds $I$ bits of
information, then during any Landauer erasure, the average amount
of heat that must be dumped to the bath is at least $I k_B T \ln 2
$ where $k_B$ is Boltzmann's constant and $T$ is the temperature
of the bath \cite{landauer}.

The principle lies at the heart of the resolution of Maxwell's
demon paradox \cite{bennett,penrose,md2}, an important factor that
stimulates continued interest on the thermodynamics of information
processing. The demon has also been investigated quantum
mechanically \cite{zurek,lloyd,bender,md2}. Also, several
alternative proofs \cite{proof1,proof2,jacobs,yu} and
generalizations \cite{maroney-abs,maroney-gen,turgut} of LEP have
been given and different approaches aimed at quantum information
processing have been developed \cite{anderson,sagawa}. The crucial
ingredient in all alternative approaches to LEP is the description
of the operation on the device as a unitary transformation on a
larger combined system of the device and the bath where initially
the two are uncorrelated and the bath is in thermal
equilibrium\footnote{For classical devices, the analogous feature
of the transformation on the combined system is the fact that this
is a canonical map, which preserves phase-space volumes by
Liouville's theorem. LEP can be deduced by following a similar
line of reasoning.}. Unitarity of this transformation is an
absolute necessity for all physical operations \cite{oqs} while
the noncorrelation and equilibrium assumptions are well-suited for
all possible experiments. Although this viewpoint is contested
\cite{norton}, LEP directly follows from the unitarity: in a
Landauer erasure, the transformation maps different initial
device-states to the same final device-state. Since the unitary
transformation has to be one-to-one, this can only happen if the
state of the bath is also modified. Finally, as the bath was
initially in equilibrium, changes in its state imply an increase
in its energy. Exact calculations then show that the average
increase in the energy of the bath must exceed the Landauer bound.

Can there be further implications of the unitarity of the transformation
on the energy exchange with the bath? A recent article gives an affirmative answer
to this question for
the case of processes that manipulate classical information \cite{turgut}. That article presents an
alternative formulation of LEP in terms of the detailed heat
transfers into the bath. In this non-entropic approach, one takes
the individual heats dumped to the bath for each initial logical
state as the main quantities of interest and places restrictions
between them. For example, if $q_i$ is the heat dumped to the bath
during a Landauer erasure when the initial logical state is $i$,
then it can be shown that
\begin{equation}
  \sum_i e^{-q_i/k_B T} \leq 1\quad,
  \label{eq:classical_landauer_exp}
\end{equation}
an inequality which is first derived by Szilard \cite{szilard}.
The main advantage of this formulation is its independence of the
coding of the information into the logical states of the device.
The standard statement of LEP, however, is dependent on the
coding: if the device is used for storing information in such a
way that the logical state $i$ is used with frequency $p_i$, then
the device has information storage capacity of $I=-\sum_i
p_i\log_2 p_i$ bits and the Landauer bound states that
\begin{equation}
  \bar{q} \geq Ik_BT\ln2
  \label{eq:classical_landauer_avg}
\end{equation}
where $\bar{q}=\sum_i p_i q_i$ is the average heat dumped to the
bath when the information in the device is erased.

Note that Eq.~(\ref{eq:classical_landauer_avg}) is valid for any
distribution $p$, hence it provides an infinite number of
restrictions on $q_i$. It can be shown that
Eq.~(\ref{eq:classical_landauer_avg}) holds for all possible
distributions $p$ if and only if
Eq.~(\ref{eq:classical_landauer_exp}) is satisfied. Thus,
Eqs.~(\ref{eq:classical_landauer_exp}) and
(\ref{eq:classical_landauer_avg}) are two equivalent formulations
of LEP. The standard form of LEP,
Eq.~(\ref{eq:classical_landauer_avg}) is more profound as it
directly connects two disparate disciplines, thermodynamics and
information theory. Yet, Eq.~(\ref{eq:classical_landauer_exp}) is
also somewhat attractive as it is a concise inequality that
captures all of the infinitely many inequalities in
Eq.~(\ref{eq:classical_landauer_avg}). Both inequalities can be
generalized to arbitrary nondeterministic processes. It is
observed that the generalization of
Eq.~(\ref{eq:classical_landauer_exp}) contains further
restrictions on heat exchanges that cannot be derived from the
entropic restrictions of the generalized LEP \cite{turgut}.

The purpose of this article is to adapt that coding-independent
approach to quantum operations applied on devices holding quantum
information. The major difference between the classical and
quantum cases is in the description of the state: instead of a
discrete index $i$, we now have a density matrix that describes
the state of the device. Hence, in place of the numbers $q_i$, we
now have a hermitian operator $Q$ whose expectation value with the
density matrix gives the average heat dumped to the bath. This
operator will be called as the \emph{heat transfer operator}
(HTO).

The unitary transformation 
can be thought as moving information around
in the combined system of the device and the bath. The resulting
information exchange between the device and the bath changes the
state and hence the energy of the bath. The HTO essentially
provides a measure of this information exchange in terms of the
energy given to the bath. The nature of the operation applied on
the device (e.g., whether the information in the device is erased
completely or only partially, or whether randomness from the bath
is introduced to the device or not) has a strong effect on the
HTO. This article investigates that relation between quantum
operations and the associated HTOs. This relation essentially
follows from the unitarity of the transformation on the combined
system. However, it appears that the unitarity property is
unnecessarily too restrictive; all of the results in this article
can also be obtained from a weaker assumption, namely that the
transformation is an isometry. This assumption is made throughout
the article.

The quantum operation-HTO relation is a generalization of the
classical LEP to the quantum domain and hence it is primarily of
theoretical interest. But, it is possible that this may find some
applications as well. For example, the restrictions on HTOs can be
used as a way to theoretically quantify the ``information erasing
nature'' or some other aspect of generic quantum operations. In
addition to this, the current work can also be useful in the
design of non-unitary operations (e.g., erasures) on technological
implementations of quantum devices. If the heat emitted during the
operation is desired to be minimized, it is necessary to select or
adjust the bath levels and the device-bath interaction in a
suitable way. The explicit construction of the erasure operations
that appear in this article is appropriate for that optimization
task and hence it may be a valuable guide for this purpose.

The organization of the article is as follows. In
Sec.~\ref{sec:hto}, the HTO associated with a realization of a
given quantum operation is defined. The nature of the HTO and some
of its fundamental properties are also discussed in this section.
Sec.~\ref{sec:Restrictions} is devoted to the derivation of
various restrictions on the HTOs. The first one of these is the
entropic bound of the generalized LEP, which can be expressed for
an arbitrary quantum operation. This bound is a good starting
point for the investigation of the coding-independent
characterization of quantum operation-HTO relation. After that,
quantum operations that completely erase the information in the
device are taken up and the associated HTOs are described. These
results are then used for the description of HTOs associated with
extremal quantum operations. In Sec.~\ref{sec:differences}, some
examples are given for quantum operations that do not completely
erase the information content of the device and it is shown that
LEP is not sufficient for a complete characterization of HTOs
associated with these operations. This indicates the usefulness of
the current approach. Finally, Sec.~\ref{sec:conclusions} contains
brief conclusions.

\section{Heat Transfer Operator}
\label{sec:hto}

Let A be a quantum system with a finite number of levels which is
used as a memory device for holding quantum information. A quantum
operation on this system is a trace-preserving completely-positive
map $\mathcal{E}$ that takes the initial state $\rho_A$ of the
device to a final state $\mathcal{E}(\rho_A)$. Although all
quantum operations are within the scope of this article, special
attention will be given to those that completely erase all
information in the device. The operation $\mathcal{E}$ will be
called a \emph{complete erasure} if the final state is a constant
mixed state $\rho_0$ for all initial states, i.e.,
\begin{equation}
  \mathcal{E}(\rho)=\rho_0\quad\textrm{for~all~}\rho~~.
\end{equation}
It will be called a \emph{Landauer erasure} if it is a complete
erasure with a pure final state, i.e.,
$\rho_0=\ket{\psi_0}\bra{\psi_0}$ for some $\ket{\psi_0}$.

Any quantum operation $\mathcal{E}$ on A can be realized as a unitary
transformation or an isometry $U_{AB}$ on a larger system AB,
where B denotes an additional system that
will henceforth be called as the bath. It is assumed that the device and
the bath are initially uncorrelated. Hence, if the initial state of the
bath is $\rho_B$, the initial state of the composite system is the product
state $\rho_{AB}=\rho_A\otimes\rho_B$. The final states after the
application of the isometry will usually be denoted by primes, e.g.,
$\rho_{AB}^\prime=U_{AB}(\rho_A\otimes \rho_B) U_{AB}^\dagger$. We will
say that the bath B, the state $\rho_B$ and the isometry $U_{AB}$ form a
\emph{realization} of the operation $\mathcal{E}$ if
\begin{equation}
  \mathcal{E}(\rho_A)=\rho_A^\prime=\tr_B \left( U_{AB}(\rho_A\otimes \rho_B)U_{AB}^\dagger\right)
  \label{eq:realization_def}
\end{equation}
is satisfied for all $\rho_A$, where here $\tr_B$ denotes the
partial trace over B.


It is assumed that the transformation $U_{AB}$ is obtained from
the evolution of the composite system AB with a time-dependent
Hamiltonian $H_{AB}(t)$. The basic assumption about the
independence of the process (i.e., the Hamiltonian and therefore
the transformation $U_{AB}$) from the information stored in the
device is also made. In particular this means that no measurements
are taken on the composite system AB during the process. Hence,
during the transformation, any information exchange with A occurs
entirely between A and B only and this process is described with
the Hamiltonian $H_{AB}(t)$. No third system is involved in the
transformation except in providing the energy for the necessary
work done on AB.


For discussing the energetics of the aforementioned
transformation, the time-dependent Hamiltonian $H_{AB}(t)$ of the
composite system AB has to be discussed. The transformation
process is assumed to be carried out over a finite time interval,
say between times $t_i$ and $t_f$, during which subsystems A and B
couple and $H_{AB}(t)$ is time dependent. Outside this time
interval, however, it is assumed that the two subsystems are
uncoupled and have time-independent Hamiltonians. Thus, we have
\begin{eqnarray}
  H_{AB}(t)  &=& H_A + H_B \quad\textrm{for}~t < t_i~\textrm{and} \label{eq:Hamiltonian_before} \\
  H_{AB}(t)  &=& H_A^\prime + H_B \quad\textrm{for}~t > t_f~, \label{eq:Hamiltonian_after}
\end{eqnarray}
where $H_A$ and $H_A^\prime$ are the Hamiltonians of the device
before and after the transformation respectively and $H_B$ is the
bath Hamiltonian. The Hamiltonian of the device is not relevant to
the main subject of the current article and hence there are no
restrictions on $H_A$ and $H_A^\prime$. However, initial and final
Hamiltonians of the bath are taken to be identical. This is also
operationally reasonable for realistic quantum operations where
the environment corresponds to B; a quantum operation on a device
may perhaps structurally change the device (so that initial and
final Hamiltonians may differ) but it merely couples with the
environment without changing it structurally. Finally, no
assumptions are made about the strength of the coupling and the
Hamiltonian $H_{AB}(t)$ during the time interval $t_i\leq t\leq t_f$.

The transformation $U_{AB}$ is the time-development operator
corresponding to the Hamiltonian $H_{AB}(t)$ between two times
$t_1$ and $t_2$ where $t_1\leq t_i$ and $t_f\leq t_2$. As
indicated in Eqs.~(\ref{eq:Hamiltonian_before}) and
(\ref{eq:Hamiltonian_after}), it is assumed that A and B do not
interact with each other outside the process time interval,
$(t_i,t_f)$. For this reason, no special values are needed to be
assumed about $t_{1,2}$. However, the results of the current
article can also be extended to cases where there is a weak
coupling between A and B outside the process time interval. In
this case, $t_2-t_f$ should be greater than the
thermal-equilibration time scales of the bath.


Note that for all realistic processes the operator $U_{AB}$ is
unitary, i.e., it satisfies $U_{AB}^\dagger
U_{AB}=U_{AB}U_{AB}^\dagger=\mathds{1}_{AB}$. However, for the
derivation of the results in this article, this requirement
appears to be unnecessarily too restrictive; we only need to
assume that $U_{AB}$ is an isometry, i.e., only $U_{AB}^\dagger
U_{AB}=\mathds{1}_{AB}$ is assumed to be satisfied. Therefore, we
allow the possibility that the transformation $U_{AB}$ is not
surjective. The generalization of unitary transformations to
isometries also serves another purpose. There are some quantum
operations, like Landauer erasures, which can only be realized by
proper isometries instead of unitaries. This is quite obvious for
Landauer erasures since all possible accessible final states do
not span the whole of the Hilbert space of AB, because any initial
pure state of AB is mapped to a final product state where A is in
$\ket{\psi_0}_A$ state\footnote{In this work, it is required that
the bath is initially in a canonical equilibrium state. Therefore
all pure states of B are possible as initial states and the map
$U_{AB}$ should send all of these to a state where A is in
$\ket{\psi_0}$ state.}.
As $U_{AB}$ is not surjective, it cannot possibly be unitary. Such
proper isometries can in principle be realized by a Hamiltonian
evolution, provided that ideal tools like perfectly impenetrable
barriers are used for making parts of the Hilbert space
inaccessible.


It is assumed that the bath is in canonical thermal
equilibrium state before the transformation and thus
\begin{equation}
  \rho_B = \frac{1}{Z_B}\exp\left(-\beta H_B\right)\quad,
\end{equation}
where $\beta=1/k_B T$ is the inverse temperature and $Z_B$ is the
corresponding partition function. It is appropriate to also assume
that the bath is sufficiently small so that $\rho_B$ is
manageable. We will say that B is a \emph{finite} system if $Z_B$
is finite for all temperatures ($Z_B\neq\infty$). In this case, B
has finite thermodynamical properties. Note that a single Hydrogen
atom or a single free particle placed in an infinite space has an
infinite partition function and therefore these are not finite
systems by the current definition. But, if the atom or the
particle is placed inside a large but finite box, then they will
become finite systems. Note also that any realistic system, which
has a finite number of particles constrained inside a finite
volume, will be classified as finite by this definition. Infinite
systems have individually vanishing level occupation probabilities
and consequently the analysis of such systems present some
difficulties. This is the main reason for using finite systems in
this article.

Let $E_B=\tr\,\rho_BH_B$ be the equilibrium energy of the bath.
Since there is no structural change in the bath, the heat dumped
to the bath during the realization is equal to the change in the
average energy, $\Delta E_B$, between times $t_1$ and $t_2$. It
can be expressed as
\begin{eqnarray}
  \Delta E_B &=& \tr (\rho_B^\prime - \rho_B) H_B \quad,
            \label{eq:Delta_EB_defined} \\
            &=& \tr_{AB}  \left(\rho_A\otimes \rho_B\right) \left(U_{AB}^\dagger(\mathds{1}_A\otimes H_B )U_{AB}\right)
                -E_B \nonumber
                \\
            &=& \tr_A (\rho_A Q)\nonumber
\end{eqnarray}
where $Q$ is the HTO of this realization, which is defined as
\begin{equation}
  Q = \tr_B \left[ \left(\mathds{1}_A\otimes \rho_B\right) U_{AB}^\dagger(\mathds{1}_A\otimes H_B )U_{AB} \right]
     -  E_B \mathds{1}_A    ~.
  \label{eq:Q_definition}
\end{equation}
A number of comments have to be made in order to comprehend the
real nature of the HTO. First, $Q$ is a hermitian operator defined
on the state space $\mathcal{H}_A$ of the device A. Its
expectation value is evaluated with the initial state,  $\rho_A$,
of the device. In other words, it is an operator that can be used
for expressing the energy transferred to B in terms of the state
of A. Second, only the average of $Q$ can be given a physical
meaning. It has been shown that no quantum observable can be
defined for the work done or the heat transferred during a process
that captures the detailed distribution of these
quantities\cite{hanggi}. However, if only the average $\Delta E_B$
is needed, then Eq.~(\ref{eq:Delta_EB_defined}) is the correct
definition, which eventually yields the HTO $Q$. On the other
hand, $Q$ cannot be used for computing the higher moments of the
energy transferred to B. For example, even though the average
energy transferred to B is given by $\tr\,\rho_AQ$, the quantity
$\tr\,\rho_AQ^2$ or the averages of other nonlinear functions of
$Q$ do not have the anticipated meaning. Averages of such
nonlinear functions of energy change of B should be computed
separately following the approach in Ref.~\onlinecite{hanggi}.
This crucial property of the HTO must always be kept in mind.

Once the HTO is computed, the average total work that needs to be
done for carrying out the operation can be obtained as
\begin{equation}
  W_{\textrm{tot}}= \tr\left(\mathcal{E}(\rho_A) H_A^\prime -\rho_A H_A+\rho_A  Q\right)~.
\end{equation}
As a result, for investigating the average energies transferred
during the realization of the operation $\mathcal{E}$, the only
nontrivial quantity to be determined is the heat dumped to the
bath. This is the main reason why this article concentrates solely
on the HTO.

The isometry condition $U_{AB}^\dagger U_{AB}=\mathds{1}_{AB}$ and
the fact that the bath is in thermal equilibrium imposes some
restrictions on the operator $Q$. The investigation of these
restrictions is the primary subject of this article. The main
question that needs to be answered is the following: for any given
quantum operation $\mathcal{E}$, hermitian operator $Q$ and
temperature value $T$, is it possible to find a bath B (namely a
Hamiltonian $H_B$) and an isometry $U_{AB}$ such that both
Eq.~(\ref{eq:realization_def}) and Eq.~(\ref{eq:Q_definition}) are
satisfied? We will say that $Q$ is a possible HTO for
$\mathcal{E}$ when such a bath and isometry exist. Our main
problem is then to find all necessary and sufficient conditions
for $Q$ to be a possible HTO for $\mathcal{E}$ at temperature $T$.
These conditions will enable us to discuss the thermodynamics of
quantum information processing (for given $\mathcal{E}$) without
specifying anything about the bath, the bath-device coupling and
the specific way the quantum operation is realized.

Before embarking on the investigation of this problem, it will be
convenient to list the following basic properties of the HTOs.

(i) First, note that any restriction that the HTOs obey can be
expressed entirely in terms of $\mathcal{E}$ and the ratio
$Q/k_BT=\beta Q$. To see this, note that both $\mathcal{E}$ and
$\beta Q$ can be expressed exclusively in terms of $U_{AB}$ and
$\rho_B$, with no explicit $T$ (or $H_B$) dependence, i.e.,
Eq.~(\ref{eq:realization_def}) and
\begin{eqnarray}
  \beta Q &=& - \tr_B \left[ \left(\mathds{1}_A\otimes \rho_B\right) U_{AB}^\dagger(\mathds{1}_A\otimes \ln\rho_B )U_{AB} \right]
            \nonumber \\
      & & \quad +  \mathds{1}_A\tr_B(\rho_B\ln\rho_B)\quad.
      \label{eq:Q_definition_alternative}
\end{eqnarray}
Hence, the main problem becomes: for given $\mathcal{E}$ and
$\beta Q$, can we find an appropriate $\rho_B$ and $U_{AB}$ such
that Eqs.~(\ref{eq:realization_def}) and
(\ref{eq:Q_definition_alternative}) are satisfied?


(ii) If $Q$ is a possible HTO for an operation $\mathcal{E}$,
then for any positive number $a$, the operator
$\tilde{Q}=Q+a\mathds{1}_A$ is also a possible HTO for the same
operation. In other words, it is always possible to dump an
additional heat $a$ whose amount is independent of the state of the device.

To show this, suppose that $U_{AB}$ on bath B is a realization of
$\mathcal{E}$ that yields the HTO $Q$. Let C be a two-level
system with Hamiltonian $H_C=\Delta \sigma_z/2$ where $\sigma_z$ denotes the associated
Pauli spin matrix. Now, take BC as the new bath and apply the isometry
$\tilde{U}_{ABC}=U_{AB}\otimes (\sigma_x)_C$ to the composite system ABC.
Since the transformation that C is subjected to is independent of how AB
is changed, it is obvious that $\tilde{U}_{ABC}$ realizes the same operation $\mathcal{E}$.
But, the heat dumped to the bath BC is now
\begin{equation}
  \Delta E_{BC} = \tr (\rho_AQ)+\Delta E_C\quad,
\end{equation}
where $\Delta E_C=\Delta\tanh(\Delta/2k_BT)$. If the value of $\Delta$ is
chosen such that $\Delta E_C=a$, then we get $\Delta
E_{BC}=\tr\rho_A\tilde{Q}$. This shows that $\tilde{Q}$ is also a possible
HTO for the same operation.


(iii) The set of two-tuples $(\mathcal{E},Q)$ formed from quantum
operations $\mathcal{E}$ and allowed HTOs $Q$ is a convex set. In other
words, if $Q_i$ is a possible HTO for the operation $\mathcal{E}_i$ for
$i=1,2,\ldots,n$, then for any $\lambda_i\geq0$ with $\sum_i\lambda_i=1$,
the operator $Q=\sum_i\lambda_iQ_i$ is a possible HTO for the operation
$\mathcal{E}=\sum_i\lambda_i\mathcal{E}_i$.

Before showing this, first note that realizations for each $\mathcal{E}_i$
can be assumed to involve the same bath B. This can always be done by
taking the bath sufficiently large. As a result, without loss of
generality, suppose that $U_{AB}^{(i)}$ is an isometry on AB that realizes
$\mathcal{E}_i$ and yields the associated HTO $Q_i$. Let C be an $n$-level
quantum system and let $\{\ket{1},\cdots,\ket{n}\}$ be an orthonormal
basis in the Hilbert space of C. Let the initial state of C be
$\rho_C=\sum_i \lambda_i \ket{i}\bra{i}$. Now, consider the isometry
$V_{ABC}$ on ABC given by
\begin{equation}
  V_{ABC}=\sum_{i=1}^n U_{AB}^{(i)} \otimes \left(\ket{i}\bra{i}\right)_C
\end{equation}
It is a simple exercise to show that this isometry gives rise to
the desired convex combinations, i.e., $V_{ABC}$ realizes the
operation $\mathcal{E}=\sum_i\lambda_i\mathcal{E}_i$ and the
associated HTO is $Q=\sum_i\lambda_iQ_i$. Note that, in here, C
acts like a \emph{controller} system for deciding which operation
should be applied on AB. The decision is given probabilistically
such that $\mathcal{E}_i$ is applied with probability $\lambda_i$.
Note also that the average energy of C does not change during this
process. In such a case, the overall operation and the associated
HTO are given by the expected convex combination expressions.

(iv) As a special case of the convexity result in (iii), the set of
allowed HTOs associated with a fixed quantum operation $\mathcal{E}$ is
also convex. This set is not bounded from above by (ii); but it is bounded
from below. We are primarily interested in finding lower bounds, if
possible tight ones, for this set.

\section{Restrictions on Heat Transfer Operators}
\label{sec:Restrictions}

\subsection{Entropic Restriction of LEP}

Before investigating the relations between HTOs and quantum
operations, it is appropriate to start with the state-dependent
entropic restriction of LEP. Let $S(\rho)=-\tr\,\rho\ln\rho$
denote the von Neumann entropy of the density matrix $\rho$. For
an arbitrary quantum operation, the average heat emitted to the
bath is bounded from below by the drop in the von Neumann entropy
of the state of the device.

\begin{theorem}[Generalized LEP]
\label{thm:genlep}
If $Q$ is a possible HTO for an operation $\mathcal{E}$, then
\begin{equation}
  \tr\,\rho_A Q \geq k_B T \Big[S(\rho_A)-S(\mathcal{E}(\rho_A))\Big]\quad,
\label{eq:entropic_LEP}
\end{equation}
for all states $\rho_A$ of the device A. For a given $\rho_A$, the
equality sign in (\ref{eq:entropic_LEP}) holds if and only if the
relation $\rho_{AB}^\prime = \mathcal{E}(\rho_A)\otimes \rho_B$ is
satisfied for some (and therefore for all) realization that yields
$\mathcal{E}$ and $Q$.
\end{theorem}

Note that the inequality above essentially gives infinitely many
bounds on the HTO, i.e., there is a separate bound on $Q$ for each
possible device state $\rho_A$. The generalized LEP has been
discussed in several
works\cite{landauer,penrose,maroney-abs,maroney-gen,turgut},
especially in the context of classical information processing.
However, for the purpose of treating the equality case, the
following straightforward derivation is included in here. It can
be deduced simply from the following wonderful identity for
deviations from canonical thermal equilibrium states
\begin{equation}
  \Delta E_B = k_B T \left( S(\rho_B^\prime\vert\vert\rho_B) +S(\rho_B^\prime)-S(\rho_B)\right)
\label{eq:magic}
\end{equation}
where
\begin{equation}
  S(\sigma\vert\vert\rho)=\tr\,\sigma(\ln\sigma-\ln\rho)
\end{equation}
denotes the relative entropy function. Only two inequalities will
be used for converting this identity to the relation in
Eq.~(\ref{eq:entropic_LEP}). The first one is the nonnegativity of
the relative entropy: $S(\sigma\vert\vert \rho)\geq 0$ with
equality holding if and only if $\sigma=\rho$. The second
inequality is the subadditivity of the von Neumann entropy for the
final state, i.e., $S(\rho_{AB}^\prime)\leq
S(\rho_A^\prime)+S(\rho_B^\prime)$ where the equality holds if and
only if the state is in product form
$\rho_{AB}^\prime=\rho_A^\prime\otimes\rho_B^\prime$. Finally,
using the fact that the initial and final states are related by an
isometry we conclude that $\rho_{AB}^\prime$ and $\rho_{AB}$ are
isospectral. In particular,
$S(\rho_{AB}^\prime)=S(\rho_{AB})=S(\rho_A)+S(\rho_B)$ and hence
\begin{equation}
  \Delta E_B\geq k_BT ( S(\rho_B^\prime)-S(\rho_B))\geq k_BT(S(\rho_A)-S(\rho_A^\prime))  ~,
  \nonumber 
\end{equation}
which is the desired inequality. It is easy to see that both of
the inequalities become equalities if and only if
$\rho_{AB}^\prime = \mathcal{E}(\rho_A)\otimes \rho_B$.$\Box$

Note that the equality can be satisfied only in some exceptional cases.
When that happens, we have $\rho_{AB}^\prime = \mathcal{E}(\rho_A)\otimes
\rho_B$ and therefore the state of the bath is unchanged, i.e.,
$\rho_B^\prime=\rho_B$. Consequently, the average heat dumped to bath is
$\tr\,\rho_A Q=0$. Moreover, since the initial and final density matrices
of AB are isospectral, it follows that $\rho_A$ and $\mathcal{E}(\rho_A)$
must also be isospectral.

It is of some interest to investigate the issue of whether the
bound in (\ref{eq:entropic_LEP}) can be improved by considering
the device A to be part of a larger system AX where X is not
involved in the realization. In other words, the inequality
\begin{equation}
  \tr\,\rho_A Q \geq k_BT\left(S(\rho_{AX})-S((\mathcal{E}\otimes\mathcal{I})(\rho_{AX}))\right)
  \label{eq:entropic_LEP_X}
\end{equation}
is also valid for any X and any initial state $\rho_{AX}$. Can this
generalized inequality be tighter than Eq.~(\ref{eq:entropic_LEP})? The
answer is negative. To show this, introduce another system Y such that AXY
is a purification of $\rho_{AX}$. The isometry of the realization is
applied on AB and hence the system XY is untouched. Using primes for
denoting the final states we have
$S(\rho_A)=S(\rho_{XY})=S(\rho_{XY}^\prime)$ and
$S(\rho_{AX})=S(\rho_Y)=S(\rho_Y^\prime)$. Then we invoke the following
alternative form of the strong subadditivity of the von Neumann entropy
\cite{Nielsen}
\begin{equation}
  S(\rho_{XY}^\prime)+S(\rho_{XA}^\prime) \geq S(\rho_Y^\prime) +
  S(\rho_A^\prime)
\end{equation}
to arrive at
\begin{equation}
  S(\rho_A) - S(\rho_{A}^\prime) \geq S(\rho_{AX}) -  S(\rho_{AX}^\prime)\quad.
\end{equation}
The last inequality clearly shows that the lower bound given by
Eq.~(\ref{eq:entropic_LEP_X}) is always smaller than the bound in
(\ref{eq:entropic_LEP}). Hence, the inclusion of additional
systems cannot possibly make an improvement in the entropic bounds
of LEP.

\subsection{Restrictions for Complete Erasure Operations}

Even though Theorem \ref{thm:genlep} places some restrictions on the
possible HTOs, it is possible that these may not be complete. In other
words, there might be operators $Q$ which satisfy the inequality
(\ref{eq:entropic_LEP}) for all possible initial states $\rho_A$, but it
still may not be a possible HTO. Some examples of such operators will be
discussed in Sec.~\ref{sec:conclusions}. However, it appears that the
inequalities (\ref{eq:entropic_LEP}) are sufficient when $\mathcal{E}$ is
a complete erasure operation. This subsection contains the proof of this
statement.

But, before that, a concise form of the inequalities in
(\ref{eq:entropic_LEP}) should be found and for this reason it is
necessary to introduce a real-valued function of hermitian
operators. The needed function is the Legendre transform $J$ of
the von Neumann entropy function $S$. For any hermitian $G$, this
function is defined by
\begin{equation}
  J(G) = -\ln \left(\tr\, e^{-G}\right)\quad.
\end{equation}
In order to arrive at the main property of this function, first
note that for any hermitian operator $G$, the operator
\begin{equation}
\sigma_G =\exp(J(G))e^{-G}
\end{equation}
is a density matrix. Then, using the nonnegativity of
$S(\rho\vert\vert\sigma_G)$ we find
\begin{equation}
  J(G) + S(\rho) \leq \tr\,\rho G
\end{equation}
an inequality which is valid for any density matrix $\rho$ and any
hermitian $G$. From this inequality it is possible to derive the
following Legendre transformation identities
\begin{eqnarray}
  J(G) &=& \min_\rho \left(\tr\, \rho G - S(\rho)\right)\quad,  \label{eq:Legendre_J_from_S}\\
  S(\rho) &=& \inf_G \left(\tr\, \rho G - J(G) \right)\quad,    \label{eq:Legendre_S_from_J}
\end{eqnarray}
where in Eq.~(\ref{eq:Legendre_J_from_S}) the minimization is over
all density matrices and the minimum is reached when
$\rho=\sigma_G$. In Eq.~(\ref{eq:Legendre_S_from_J}) the infimum
is over all hermitian operators $G$. If $\rho$ has no zero
eigenvalues, then right-hand side of that equation is minimized by
$G=-\ln \rho$. Otherwise, if $\rho$ has zero eigenvalues, then the
right-hand side of (\ref{eq:Legendre_S_from_J}) has no minimum;
but infimum is obtained by a sequence of density matrices which
approach to the limit $G_{\textrm{lim}}=-\ln\rho$, which is now an
operator that has some infinite eigenvalues.

Consider a complete erasure operation $\mathcal{E}$ where
$\mathcal{E}(\rho_A)=\rho_0$ for all states $\rho_A$. In this
case, the inequality (\ref{eq:entropic_LEP}), which is valid for
all density matrices $\rho_A$, can be written compactly using the
$J$-function as
\begin{equation}
  J(\beta Q) \geq -S(\rho_0)\quad.
\label{eq:Cond_Complete_Erasure}
\end{equation}
Moreover, this condition is almost sufficient for $Q$ to be a
possible HTO. There is only a slight complication in the treatment
of the equality case which is related to the finiteness property
of the bath. The following theorem gives the complete
characterization of HTOs for complete erasures.

\begin{theorem}
\label{thm:complete} Let $Q$ be a hermitian operator and $\mathcal{E}$ be
a complete erasure to the final state $\rho_0$. When only realizations
involving finite baths are considered, $Q$ is a possible HTO for
$\mathcal{E}$ if and only if
\begin{itemize}
\item[(i)] either $J(\beta Q)>-S(\rho_0)$,
\item[(ii)] or $J(\beta Q)=-S(\rho_0)$ and $\sigma_{\beta Q}$ is
isospectral with $\rho_0$.
\end{itemize}
\end{theorem}

Therefore the equality sign in (\ref{eq:Cond_Complete_Erasure})
applies only under very rare situations where the eigenvalue
spectrum of $Q$ must satisfy strict conditions; in particular the
dimension of the state space of A must be equal to the matrix rank
of the final state $\rho_0$. If this is not the case, then the
strict inequality holds.

The proof of the necessity of the conditions (i) and (ii) has
already been argued above in obtaining
Eq.~(\ref{eq:Cond_Complete_Erasure}); only a treatment of the
equality case is left. If equality holds in
(\ref{eq:Cond_Complete_Erasure}), then the density matrix that
minimizes the expression (\ref{eq:Legendre_J_from_S}), namely
$\rho_A=\sigma_{\beta Q}$, gives equality in
(\ref{eq:entropic_LEP}), and therefore by Theorem
\ref{thm:genlep}, $\rho_A$ must be isospectral with
$\mathcal{E}(\rho_A)=\rho_0$, i.e., the condition in (ii) holds.
This completes the proof of the necessity of the conditions of
Theorem \ref{thm:complete}.

For the sufficiency proof, we have to start with a given $Q$ that
satisfies either (i) and (ii) and then show that a realization
that yields $\mathcal{E}$ and $Q$ can be constructed. The needed
isometry is actually quite simple. It essentially copies the
entire information on the device to some subsystem of the bath, so
that the information is completely erased, and at the same time it
copies the thermal equilibrium state of some subsystem of B to the
device, so that the device has always the same final state. Even
though the isometry is simple, the job of adjusting the HTO to be
equal to the given operator $Q$ is technically complicated. For
this reason, the proof of the sufficiency of the conditions in
Theorem \ref{thm:complete} is done separately in Appendix
\ref{sec:app_com}. However, it is appropriate to give a separate
proof in here for the special case of Landauer erasures for which
the required bath is simpler. Treatment of the technical details
in this proof will also simplify the presentation of the full
proof of Theorem \ref{thm:complete} in Appendix \ref{sec:app_com}.

Let $\mathcal{E}$ be a Landauer erasure to the standard state
$\ket{\psi_0}$ ($\mathcal{E}(\rho_A)=\ket{\psi_0}\bra{\psi_0}$ for
all $\rho_A$). Note that the special case (ii) of Theorem
\ref{thm:complete} can hold only if the dimension of the Hilbert
space of A is $d=1$. This is a very trivial case where the system
A cannot hold any information at all. In such a case, the HTO $Q$
is a number and the theorem just says that this number is
nonnegative, $Q\geq0$. For the generic case of $d\geq2$, however,
only the strict inequality of (i) holds. The following is merely a
corollary of Theorem \ref{thm:complete} but we state it as a
separate theorem and give a separate proof.

\begin{theorem}
\label{cor:Landauer}
 Let $\mathcal{E}$ be a Landauer erasure on a device A having
$d$ levels with $d\geq2$ and let $Q$ be a hermitian operator. The
following are equivalent.
\begin{itemize}
\item[(i)] $Q$ is a possible HTO for a realization on a finite
bath.
\item[(ii)] $\tr\, e^{-\beta Q} < 1$.
\item[(iii)] $\tr\, \rho_A Q>k_BT S(\rho_A)$ for all $\rho_A$.
\end{itemize}
\end{theorem}

It is obvious that the statements (ii) and (iii) are equivalent;
both are simply stating that $J(\beta Q)$ is a strictly positive
number. The statement (iii) is the conventional way of expressing
the LEP. Statement (ii) is the quantum analog of the inequality in
Eq.~(\ref{eq:classical_landauer_exp}). (The strict inequality is a
minor mathematical detail; it arises from our insistence on using
finite baths.)

It has been shown in Theorem \ref{thm:genlep} that (i) implies (ii) and
(iii). What is left is the proof of the implication (ii)$\Rightarrow$(i).
Let $Q$ be a hermitian operator such that $J(\beta Q)>0$. A realization of
the Landauer erasure will be constructed below and it will be shown that
the associated HTO is equal to $Q$.


\begin{figure}
\includegraphics[scale=0.8]{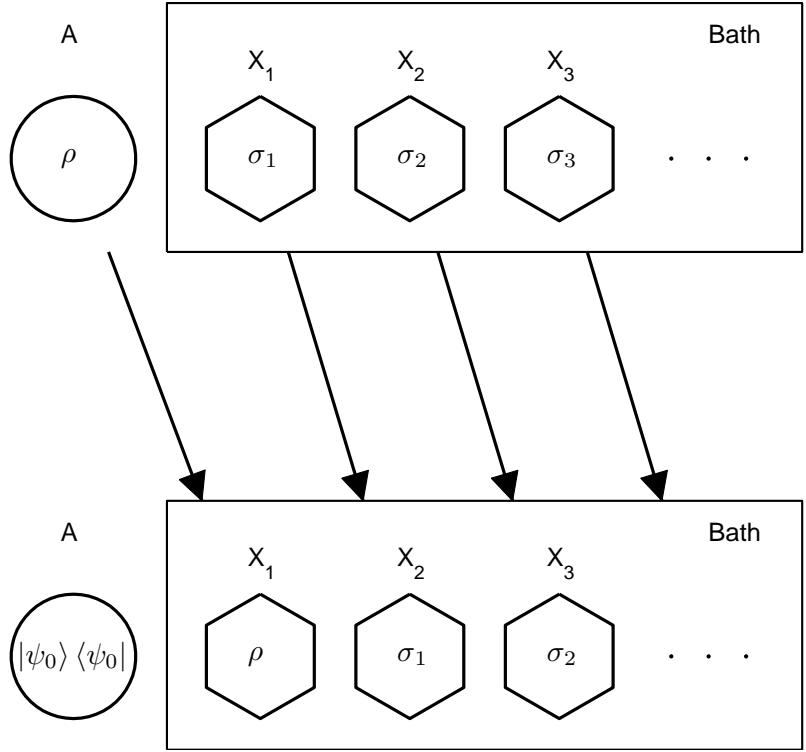}
\caption{The pictorial description of the isometry used in the
proof of Theorem \ref{thm:complete} for a Landauer erasure
operation. Here, the state of the device A is made equal to the
standard state $\ket{\psi_0}$ while the original information
content of each subsystem is shifted to the right.}
\label{fig:Landauer}
\end{figure}


\paragraph{The bath and its basis states.}
It is obvious that the bath needed for this purpose has infinitely
many levels. Only in that case, it is possible to copy the
information on A onto B and at the same time, retain the
information already present in B. For better visualizing the
information flow, we will think of the bath as being composed of
infinitely many copies of a $d$-level system X. The constitution
of B can be expressed as
$\mathrm{B}=\mathrm{X}_1\mathrm{X}_2\mathrm{X}_3\cdots$. Let the
standard orthonormal basis of X be denoted by $\{\ket{i}\}$ where
$i=0,1,\ldots,d-1$. The following self-explanatory shorthand will
be used for the state vectors of the bath
\begin{equation}
  \ket{i_1,i_2,i_3,\cdots}_B = \ket{i_1}_{X_1}\otimes\ket{i_2}_{X_2}\otimes\ket{i_3}_{X_3}\otimes\cdots~~.
\end{equation}
The Hilbert space $\mathcal{H}_B$ of the bath will be defined as
the linear span of the above states where the sequence $\{i_k\}$
contains only \emph{finitely many nonzero elements}. When the
Hamiltonian is defined below, it will be seen that the states
which are not included into $\mathcal{H}_B$ have infinite energy
and zero equilibrium-state probabilities. Consequently such states
will never enter into the discussion.

In addition to this, defining $\mathcal{H}_B$ in such a way makes
it a \emph{separable} Hilbert space having a countable orthonormal
basis. To see this, note that the sequence $\{i_k\}$ can be
considered as the digits of the base-$d$ representation of an
integer $n$, written in short as $n=(\cdots i_3i_2i_1)_d$. The
integer $n$ is given in terms of the sequence as
\begin{equation}
  n = i_1+i_2 d+i_3d^2+\cdots=\sum_{k=1}^\infty i_k d^{k-1}~~.
\end{equation}
Consequently, there is a one-to-one correspondence between
nonnegative integers $n$ and sequences $\{i_k\}$ that has finitely
many nonzero elements. For this reason, the standard orthonormal
basis of $\mathcal{H}_B$ can be enumerated by a nonnegative
integer $n$ and therefore the alternative label
\begin{equation}
 \ket{n}_B=\ket{i_1,i_2,i_3,\cdots}_B
\end{equation}
can be used. It is thus clear that the Hilbert space
$\mathcal{H}_B$ has a countable basis and hence it is separable.


\paragraph{The isometry.}
Let $q_i$ ($i=0,1,\ldots,d-1$) be the eigenvalues of $Q$ arranged
in nondecreasing order and let $\{\ket{i}\}$ be the orthonormal
basis of $\mathcal{H}_A$ formed from the corresponding
eigenvectors.
\begin{equation}
   Q = \sum_{i=0}^{d-1} q_i \ket{i}\bra{i}   \qquad \,  (q_0\leq q_1\leq \cdots\leq q_{d-1})~.
   \label{eq:Q_spectral_decomp}
\end{equation}
The isometry $U_{AB}$ on the composite system AB is defined as
follows
\begin{equation}
  U_{AB}\ket{\ell}_A \otimes \ket{i_1,i_2,i_3,\cdots}_B
    = \ket{\psi_0}_A \otimes \ket{\ell, i_1,i_2,\cdots}_B~~,
\end{equation}
for any allowed values of input labels. In words, the isometry
sets the state of A to the standard state $\ket{\psi_0}$ and
copies its contents onto the subsystem X$_1$. It also forms an
infinite chain of information flow inside the bath as the contents
of X$_k$ are copied onto X$_{k+1}$ ($k\geq1$) for all $k$. The
flow of information is pictorially depicted in
Fig.~\ref{fig:Landauer}. Essentially the isometry shifts the
information content of all subsystems of AB onto the next
subsystem on the right.

This flow is highly reminiscent of the motion of the guests in
Hilbert's Hotel Infinity paradox. In that hotel, all of the
infinitely many rooms are occupied. But, it is still possible to
accommodate a new guest by moving everyone to the next room. The
isometry $U_{AB}$ essentially does the same for information. It
retains the whole information already stored in B while it
squeezes additional information coming from A into B. Our hotel,
however, has an important difference from Hilbert's hotel: as it
follows from LEP, there is a nonzero check-in cost. In other
words, while squeezing new information into B, the average energy
of B increases. To be able to compute that increase, the bath
Hamiltonian must be specified.

\paragraph{The Hamiltonian of the bath.}
The Hamiltonian will be constructed such that each subsystem of
the bath is independent and the energy eigenstates are identical
with the standard basis states. The precise energy eigenvalues of
the subsystem X$_1$ will be chosen to be identical with those of
$Q$ (plus a possible shift), i.e., $H_{X_1}=Q+c_1\mathds{1}_{X_1}$
in the standard basis where here, $c_1$ is some constant. The
energy eigenvalues of the remaining subsystems X$_2$, X$_3$,
$\ldots$ will only shift the HTO by a multiple of identity. With
this choice, the energy change of the bath is $\Delta
E_B=\tr(\rho_AQ)+c_2$ where $c_2$ is a constant that is
independent of $\rho_A$. The value of $c_2$ depends on the energy
levels of X$_2$, X$_3$, $\ldots$. By adjusting these levels
appropriately, it is possible set $c_2=0$. This shows that the HTO
is $Q$ and completes the proof. Since the details are complicated,
the adjustment of the energy levels of other subsystems is
discussed in Appendix~\ref{sec:app_lan}. $\Box$

The realization described in the proof above can be used for
finding conditions for minimizing the average heat emission during
a Landauer erasure operation. Suppose that A is used as a memory
device in an implementation of a quantum computer. Suppose also
that A is subjected to a Landauer erasure at ambient temperature
$T_{\mathrm{env}}$ in some steps of the operation of the computer.
Let $\rho_A^{\mathrm{av}}$ denote the average state of the device
before the erasures are applied. The average heat emitted per
erasure is then $\bar{q}=\tr\,\rho_A^{\mathrm{av}}Q$ where $Q$ is
the HTO. The problem is to minimize $\bar{q}$ under the
restriction that $Q$ satisfies the conditions mentioned in Theorem
\ref{cor:Landauer}. According to Theorem \ref{thm:genlep},
$\bar{q}$ is bounded below by
$k_BT_{\mathrm{env}}S(\rho_A^{\mathrm{av}})$. A simple inspection
shows us that this bound is attained by a small error by the HTO
\begin{equation}
  Q = -k_BT_{\mathrm{env}} \ln \rho_A^{\mathrm{av}} + \epsilon\mathds{1}_A
\end{equation}
where $\epsilon$ is a small positive number. The construction in
the proof of Theorem \ref{cor:Landauer} can then be followed for
adjusting the energy levels of the bath, the detailed Hamiltonian
$H_{AB}(t)$ during device-bath interaction and the consequent
isometry. In a realistic situation where the environment serves as
the bath B, it is not possible to modify the energy levels of B at
will and the Landauer bound may not be attained; but the
construction above can still be useful. For example, a finite
chain X$_1$X$_2\cdots$X$_N$ can be used for connecting the device
and the environment. By adjusting the energy levels of the
subsystems in the chain and using the same information flow,
$\bar{q}$ can still be significantly decreased.

\subsection{General Quantum Operations}

A generic quantum operation causes only a partial erasure of the quantum
information stored by the device by forming an entanglement between the
device and the bath during the realization. Even though the inequality
(\ref{eq:entropic_LEP}) continues to be valid for the generic case, it
does not necessarily give a complete characterization of the possible HTOs
as it will be discussed in Sec. \ref{sec:conclusions}. This subsection
contains a few basic results about the HTOs of generic quantum operations
that surpass (\ref{eq:entropic_LEP}) in implications.

Let $\mathcal{E}$ be a quantum operation on the device A. Such an operation
can be written in the Kraus form
\begin{equation}
  \mathcal{E}(\rho_A) = \sum_{i=1}^n M_i \rho_A M_i^\dagger
  \label{eq:Kraus}
\end{equation}
where the Kraus operators $M_i$ satisfy
\begin{equation}
  \sum_{i=1}^n M_i^\dagger M_i = \mathds{1}_A\quad.
\end{equation}
Kraus operators can be chosen in various different
ways\cite{Nielsen}. However, a representation that has the
smallest number $n$ of terms will be preferred in this article. In
this case, the set of Kraus operators $\{M_1,\ldots,M_n\}$ are
linearly independent.

Now, let $U_{AB}$ be the isometry for a realization of the
operation $\mathcal{E}$. This isometry can be written as
\begin{equation}
  U_{AB} = \sum_{i=1}^n (M_i)_A \otimes (L_i)_B
  \label{eq:U_AB_expansion}
\end{equation}
for some operators $L_i$ acting on the Hilbert space of the bath
B. This can be shown by completing the set of operators
$\{M_1,M_2,\ldots,M_n\}$ to a basis for the operator space by
adding new, linearly-independent operators
$M_{n+1},M_{n+2},\cdots$ to the list, and then applying the
expansion in Eq.~(\ref{eq:U_AB_expansion}) with a larger upper
limit of summation. As $U_{AB}$ is a realization of $\mathcal{E}$,
we have
\begin{eqnarray}
  \mathcal{E}(\rho_A) &=&  \tr_B U_{AB}(\rho_A\otimes  \rho_B)U_{AB}^\dagger \\
                &=& \sum_{ij} M_i \rho_A M_j^\dagger \left( \tr\,\rho_BL_j^\dagger L_i\right).
\end{eqnarray}
The linear independence of $\{M_i\}$ implies the linear independence of
the set of maps $\{M_i \rho_A M_j^\dagger\}_{i,j}$ on the density
matrices. For this reason, we have
\begin{equation}
  \tr\,\rho_BL_j^\dagger L_i = \left\{
     \begin{array}{ll}
        \delta_{ij}  & \textrm{for}~i,j \leq n \\
        0            & \textrm{for}~i>n~\textrm{or}~j> n
     \end{array}
     \right.
\end{equation}
and consequently $\tr\,\rho_B L_i^\dagger L_i=0$ for all $i>n$. Since
$\rho_B$ is a canonical equilibrium state, it has full rank and therefore
$L_i=0$ for all $i>n$. This shows that the summation in
Eq.~(\ref{eq:U_AB_expansion}) contains only $n$ terms. The following
theorem directly follows from this expansion.

\begin{theorem}
\label{thm:gen1} Let $\mathcal{E}$ be the quantum operation given
in (\ref{eq:Kraus}) where $M_i$ are linearly independent.
\begin{itemize}
\item[(a)] If $Q$ is an HTO for $\mathcal{E}$, then there is an
$n\times n$ hermitian matrix $q$, (which might be called as the
\emph{heat transfer matrix}) such that
\begin{equation}
  Q=\sum_{i,j=1}^n  q_{ij} M_i^\dagger M_j \quad.
\label{eq:Q_q_relation}
\end{equation}
\item[(b)] If $n\geq2$ and the set of $n^2$ operators
$\{M_i^\dagger M_j\}_{i,j=1}^n$ is linearly independent, then $Q$
is an HTO for $\mathcal{E}$ if and only if the unique matrix $q$
in (\ref{eq:Q_q_relation}) satisfies
\begin{equation}
  \tr\, e^{-q/k_B T} < 1~~~.
\label{eq:q_inequality}
\end{equation}
For $n=1$, $Q$ is an HTO if and only if $Q=\alpha\mathds{1}_A$
where $\alpha\geq0$.
\item[(c)] If $\{M_i^\dagger M_j\}_{i,j=1}^n$ is linearly
dependent, then the condition (\ref{eq:q_inequality}) is
sufficient for $Q$ given by (\ref{eq:Q_q_relation}) to be a
possible HTO; but that condition is not necessary.
\end{itemize}
\end{theorem}

Note that the set of possible quantum operations on A is a convex set. It
can be shown that the extreme points of this convex set correspond to
quantum operations for which the set of operators $\{M_i^\dagger
M_j\}_{i,j=1}^n$ is linearly independent\cite{BhatiaPos}. Therefore, part
(b) of the theorem above gives a complete characterization of the HTOs of
all extreme quantum operations.

\textit{Proof:}
For (a), the use of the definition of the HTO given in
Eq.~(\ref{eq:Q_definition}) directly leads to
Eq.~(\ref{eq:Q_q_relation}) where
\begin{equation}
  q_{ij} = \tr \left(\rho_B L_i^\dagger H_B  L_j\right)-E_B\delta_{ij}\quad.
\end{equation}
To show (b), first write the isometry condition $U_{AB}^\dagger
U_{AB}=\mathds{1}_{AB}$ as
\begin{equation}
    \sum_{i,j=1}^n (M_i^\dagger M_j)_A \otimes (L_i^\dagger L_j)_B =\mathds{1}_A\otimes \mathds{1}_B~~.
\end{equation}
The linear independence of $\{M_i^\dagger M_j\}_{i,j=1}^n$ then leads to
the following relations
\begin{equation}
  L_i^\dagger L_j = \delta_{ij} \mathds{1}_B\quad.
\end{equation}
In other words, each $L_i$ is an isometry on the bath that maps
$\mathcal{H}_B$ onto mutually perpendicular subspaces of $\mathcal{H}_B$.
The last set of relations is actually equivalent to the statement that
$U_{AB}$ is an isometry.

Consider now a hypothetical $n$-level system R. Let $\{\ket{i}\}_{i=1}^n$
be an orthonormal basis for the Hilbert space of R. The following operator
on the composite system RB,
\begin{equation}
  W_{RB} = \sum_{i=1}^n (\ket{1}\bra{i})_R \otimes (L_i)_B
  \label{eq:W_RB}
\end{equation}
is then an isometry. It can also be readily verified that $W_{RB}$
is an isometry if and only if $U_{AB}$ is also an isometry. The
most important point in here is that $W_{RB}$ is a realization of
a Landauer erasure on R. Furthermore, the HTO associated with this
erasure is given by
\begin{eqnarray}
    Q^\prime_R &=& \tr_B \left[ \left(\mathds{1}_R\otimes \rho_B\right) W_{RB}^\dagger(\mathds{1}_R\otimes H_B )W_{RB} \right]
               -  E_B \mathds{1}_R\quad. \\
      &=& \sum_{i,j} q_{ij} \ket{i}\bra{j}~~.
     \label{eq:Qprime_R}
\end{eqnarray}
For $n\geq2$, the application of Theorem \ref{cor:Landauer} proves
the claim. The result for the special case of $n=1$ also follows
trivially from the discussion in this subsection.

Part (c) follows from the converse of the argument used above for
part (b). Suppose that $q$ satisfies (\ref{eq:q_inequality}). In
that case, $Q^\prime_R$ given by Eq.~(\ref{eq:Qprime_R}) is an
allowed HTO for a Landauer erasure. Hence one can construct the
associated realization; let $W_{RB}$ be the isometry for this
realization. We use Eq.~(\ref{eq:W_RB}) to define $L_i$ and then
use Eq.~(\ref{eq:U_AB_expansion}) to define $U_{AB}$. It is
straightforward to verify that $U_{AB}$ is the desired isometry
that realizes $\mathcal{E}$ and has $q$ as the heat transfer
matrix.

Finally note that any $q$ that satisfies (\ref{eq:q_inequality})
necessarily yields a positive definite $Q$. However, Theorem
\ref{thm:complete} allows HTOs that may have negative eigenvalues. This
shows that the condition (\ref{eq:q_inequality}) is not necessary. The
proof of Theorem \ref{thm:gen1} is thus completed. $\Box$

Note that part (b) of Theorem \ref{thm:gen1} is proved by finding
a closely related Landauer erasure on a hypothetical system. It is
possible to extend the scope of part (b) by making use of complete
erasures. However, the expressions used in such an extension do
not appear to be particularly illuminating and hence they are not
given in here. Theorem \ref{thm:gen1} is a good starting point for
discussing the properties of HTOs associated with generic quantum
operations. The following result is an example. It is essentially
a different way of saying that the set of HTOs associated with a
given quantum operation has no upper limit.

\begin{theorem}
Let $q$ be a heat transfer matrix for $\mathcal{E}$ which is
related to the corresponding HTO by (\ref{eq:Q_q_relation}). Then,
for any $q^\prime$ with $q^\prime>q$, $q^\prime$ is also a
possible heat transfer matrix.
\end{theorem}

To prove this, first define the matrix $s=q^\prime-q$ and observe that it
is strictly positive definite, i.e., all eigenvalues of $s$ are positive.
In such a case, it is possible to find a positive number $t\geq1$ such
that the matrix $q+ts$ satisfies (\ref{eq:q_inequality}) and therefore
$q+ts$ is a possible heat transfer matrix by part (c) of Theorem
\ref{thm:gen1}. Finally, by the convexity of the set of HTOs, the matrix
\begin{equation}
  q^\prime = \frac{1}{t}(q+ts)+\frac{t-1}{t}q
\end{equation}
is a possible heat transfer matrix. This completes the proof.
$\Box$

Note that this theorem cannot be extended by replacing the
matrices $q$ by the associated HTOs mainly due to the restriction
in Theorem \ref{thm:gen1}~(a) which must be satisfied by all HTOs.
It is plausible that the strict inequality $q^\prime>q$ can be
replaced by the weaker requirement of $q^\prime\geq q$, but this
conjecture is still awaiting proof.

\section{Differences from Landauer's Bound}
\label{sec:differences}

The restrictions satisfied by the HTOs are closely related to LEP since
both follow only from the condition that the transformation on the
composite system is an isometry. However, LEP only relates the dumped heat
to the drop in the von Neumann entropy of the state of the device. For
complete erasures, it turned out that LEP also gives a complete
characterization of the HTOs. In other words, Theorems \ref{thm:complete}
and \ref{cor:Landauer} are identical in content with Theorem
\ref{thm:genlep}; they are just expressing the same statement using
different equations. As it will be shown below, this is not the case for
other quantum operations, i.e., it is possible to find examples of
operations $\mathcal{E}$ that are not complete erasures, such that the
HTOs associated with $\mathcal{E}$ satisfy further restrictions that
cannot be explained by LEP alone. Such operations can be considered to be
``partial erasures'' as some part of the initial information present in
the device will be retained (since $\mathcal{E}(\rho)$ depends on $\rho$).
The Landauer bound in Theorem \ref{thm:genlep} still applies to such
$\mathcal{E}$, but this bound fails to imply the additional features of
HTOs that are stated in Theorem \ref{thm:gen1}.

For example, the special structure of the HTOs given in Theorem
\ref{thm:gen1}~(a) does not appear in the generalized LEP inequality
(\ref{eq:entropic_LEP}). To see this clearly, consider unitary rotations
(or isometries) as quantum operations, i.e., $\mathcal{E}(\rho_A)=W\rho_A
W^\dagger$ where $W^\dagger W=\mathds{1}_A$. In such a case, the entropic
bound (\ref{eq:entropic_LEP}) merely states that the HTO must be a
positive-semidefinite operator. However, in reality, the HTOs should not
only be positive-semidefinite but they should also be a multiple of
identity as asserted in Theorem \ref{thm:gen1}~(b). Only in such a case,
the average heat dumped to the bath becomes independent of the state of
the device. Otherwise, any $Q$ which is not a multiple of identity would
create serious problems for quantum mechanics: it implies that by
investigating the energy of the bath, some information can be obtained
about the state of A without disturbing it! It is apparent that Theorem
\ref{thm:gen1}~(a) is intimately related with the characteristic features
of quantum information.

Apart from this, there are situations where a given $Q$ satisfies the
special form given in Theorem \ref{thm:gen1}~(a) and the entropic
inequalities (\ref{eq:entropic_LEP}), but it still fails to be an HTO. As
an example, consider the special problem of finding an HTO of the form
$Q=\alpha\mathds{1}_A$ in such a way that $\alpha$ has the minimum
possible value. If the quantum operation is extremal, then by Theorem
\ref{thm:gen1}~(b), $Q$ is an HTO if and only if $\alpha>k_BT\ln n$ where
$n$ is the minimum number of Kraus operators (for $n\geq2$). However, if
only the entropic inequalities (\ref{eq:entropic_LEP}) are used, it is
possible to obtain a different lower bound for $\alpha$, especially if the
quantum operation is near to identity.

As an example, consider the one-parameter family $\mathcal{E}_t$ of
quantum operations defined by $n=2$ Kraus operators
\begin{eqnarray}
  M_1(t) &=& t X\quad,\\
  M_2(t) &=& \sqrt{\mathds{1}-t^2 X^\dagger X}\quad,\\
  \mathcal{E}_t(\rho_A) &=& \sum_{i=1}^2 M_i(t)^\dagger \rho_A  M_i(t) \quad,
\end{eqnarray}
where $X$ is a non-hermitian operator such that $\{M_i^\dagger(t)
M_j(t)\}_{i,j=1}^2$ are linearly independent. Note that in the limit
$t\longrightarrow0$, the operation $\mathcal{E}_t$ approaches to the
identity transformation. Therefore, the maximum drop in the von Neumann
entropy of the device,
\begin{equation}
  B_t = \max_{\rho_A} \left( S(\rho_A)-S(\mathcal{E}_t(\rho_A))  \right)\quad,
\end{equation}
approaches zero in the same limit. Then, the smallest possible value of
the coefficient $\alpha$ allowed by the entropic restrictions of LEP is
$\alpha_{\textrm{min}}=k_BT B_t$, which also tends to zero as
$t\longrightarrow0$. However, it is shown above that $\alpha>k_BT\ln2$,
independent of how close $\mathcal{E}_t$ to the identity operation. This
clearly shows that Theorem \ref{thm:gen1}~(b) contains restrictions on
HTOs that are not implied by the entropic LEP restrictions.

In order to eliminate a possible misunderstanding that may arise from the
previous example, let us stress on the fact that once we lift the
restriction that $Q$ is proportional to the identity, it is possible to
find HTOs that approach to the zero operator when $\mathcal{E}_t$
approaches to the identity operation. For example, for $\mathcal{E}_t$
given above,
\begin{equation}
  Q=\sum_{i=1}^2 q_i M_i(t)^\dagger M_i(t)\quad,
\end{equation}
with $\beta q_2=t^2$ and $\beta q_1\approx \ln(1/t^2)$ yields an HTO $Q$
with
\begin{equation}
  \beta Q\approx t^2 \left(\mathds{1}_A +\ln(1/t^2) X^\dagger X\right)
\end{equation}
which converges to zero as $t\longrightarrow0$.

\section{Conclusions}
\label{sec:conclusions}

In this article, the concept of heat transfer operator is
introduced and some of its properties are discussed. The HTO is
the main quantity needed for analyzing the energetics of quantum
operations. Hence, the complete characterization of the HTOs
associated with a given operation $\mathcal{E}$ occupies a central
place in the thermodynamics of quantum information processing.
This article mainly contains results on this problem. A complete
characterization could be provided only for complete erasures and
extreme operations. The characterization of HTOs of generic
non-extremal quantum operations is still an open problem.

The set of HTOs associated with a given quantum operation might be
considered as a means of describing the disturbance caused by the
operation on the environment. This description is provided through a set
of operators that has a clear operational interpretation. This points out
the main theoretical interest for this problem. In addition to this, both
the results obtained and the techniques used in this article can be useful
in the implementation of non-unitary operations in future quantum
computers. For example, the minimization of the average heat emission
during an erasure operation requires a careful construction of the
transformation on the extended system of the device and the bath. The
constructions used in this article will be useful for such design
problems.

\section*{Acknowledgements}

The authors are grateful to N. K. Pak for stimulating discussions.


\appendix

\section{Adjusting Bath Levels in the Proof of Theorem \ref{cor:Landauer}}
\label{sec:app_lan}

It is convenient to think of the subsystem Hamiltonians to be
dependent on a continuous parameter $\nu$. For this purpose,
consider $d$ continuous functions $\epsilon_i(\nu)$
($i=0,1,\cdots,d-1$) of a real parameter $\nu\in[0,1]$, which will
be used for energy eigenvalues of subsystems. They are chosen to
satisfy the following properties:
\begin{itemize}
\item[(i)] The zeroth level has zero energy, $\epsilon_0(\nu)=0$,
for all $\nu$.
\item[(ii)] The energies of all excited levels diverge to
$+\infty$ as $\nu$ approaches to $1$
(i.e., $\lim_{\nu\rightarrow1} \epsilon_i(\nu) = \infty$ for any $i\geq1$.)
\item[(iii)] The energy spectrum at $\nu=0$ is identical with the
eigenvalues of $Q$ up to a constant shift as
$\epsilon_i(0)=q_i-q_0$ for all $i$.
\end{itemize}

Let $\{\nu_k\}$ be an infinite, increasing sequence starting from
$0$ and converging to $1$, i.e., $\nu_1=0\leq \nu_2\leq
\nu_3\leq\cdots $ and $\lim_{k\rightarrow\infty} \nu_k=1$. The
energy levels of X$_k$ will be taken as $\epsilon_i(\nu_k)$. In
other words, in terms of
\begin{equation}
  h(\nu) = \sum_{i=0}^{d-1} \epsilon_i(\nu)\ket{i}\bra{i}\quad,
\end{equation}
the Hamiltonians of individual subsystems can be expressed as
$H_{X_k}=h(\nu_k)$. The bath Hamiltonian $H_B$ is the sum of the
subsystem Hamiltonians and therefore the energy eigenvalue of the
bath state $\ket{n}_B=\ket{i_1,i_2,i_3,\cdots}_B$ is
\begin{equation}
  E_{n}= \sum_{k=1}^\infty \epsilon_{i_k}(\nu_k)~~~.
\end{equation}
Note that if $\{i_k\}$ had infinitely many nonzero entries, then
the corresponding energy would be infinite. This justifies the
definition of the Hilbert space $\mathcal{H}_B$ in the way
described above.

Let $\sigma(\nu)=\exp(-\beta h(\nu))/\zeta(\nu)$ be the
parameter-dependent density matrix and $\zeta(\nu)$ be the
corresponding partition function. The shorthand notation
$\sigma_k=\sigma(\nu_k)$ will be used for the thermal equilibrium
states of X$_k$. The partition function of the bath is then given
by
\begin{equation}
  Z_B = \prod_{k=1}^\infty \zeta(\nu_k)~~.
\label{eq:ZB_Land}
\end{equation}
Note that the convergence of this product depends only on the
large $k$ behavior of the sequence $\{\nu_k\}$. By adjusting the
way $\nu_k$ converges to $1$, it is possible obtain a finite
$Z_B$.

Let us now compute the HTO. In the following expressions, the subscripts are used for
indicating the subsystem a given density matrix applies to.
Suppose that the initial state of the device A is $\rho$. The
initial and final states of AB are given by the following
self-explanatory expressions.
\begin{eqnarray}
  \rho_{AB} &=& ~~\,\quad (~~\rho~~)_A \otimes (\sigma_1)_{X_1} \otimes  (\sigma_2)_{X_2} \otimes \cdots~,\\
  \rho_{AB}^\prime &=& \left(\ket{\psi_0}\bra{\psi_0}\right)_A \otimes (\,\rho~)_{X_1} \otimes  (\sigma_1)_{X_2} \otimes\cdots~.
\end{eqnarray}
The change in the average energy of the bath is given by
\begin{equation}
  \Delta E_B  = \tr (\rho Q) - k_BTJ + \Delta  \label{eq:DeltaEB_expr_Land}
\end{equation}
where $J=J(\beta Q)$ and
\begin{eqnarray}
   \Delta &=& k_BTJ-q_0 - \tr\,\sigma_1H_{X_1}+\sum_{k=2}^\infty \tr  (\sigma_{k-1}-\sigma_k) H_{X_k} ~~,
   \label{eq:Delta_vs_DensMat_Land}
\end{eqnarray}
which is a constant independent of the device state $\rho$. Let us
first note that, by using the identity in Eq.~(\ref{eq:magic}), it
is possible to express $\Delta$ as
\begin{equation}
  \Delta = k_BT\sum_{k=1}^\infty     S(\sigma_k\vert\vert\sigma_{k+1}) ~~,
\end{equation}
which clearly shows that $\Delta$ is a strictly positive number.
It will be argued below that the sequence $\{\nu_k\}$ can be
selected in such a way that $\Delta$ can be adjusted to be equal
to any positive number and this can be achieved with a finite
$Z_B$. This will be done by showing that $\Delta$ can be made
arbitrarily large and arbitrarily small and then allude to
continuity to prove the claim.

(1) First, note that if the parameter $\nu_2$ of X$_2$ is
increased towards $1$, the final energy of X$_2$ approaches to
infinity. Hence, in this limit $\Delta$ diverges to $\infty$.
Therefore, $\Delta$ can be made arbitrarily large. (2) In order to
make $\Delta$ small, the parameters of neighboring subsystems
should be chosen nearly equal. In other words, the increments
$\nu_{k+1}-\nu_k$ should tend to 0. In this limit, the series in
Eq.~(\ref{eq:Delta_vs_DensMat_Land}) approach to an integral,
\begin{eqnarray}
   \sum_{k=2}^\infty \tr  (\sigma_{k-1}-\sigma_k) H_{X_k}
       &\longrightarrow&
                -\int_0^1 d\nu \, \tr \left(     \frac{\partial \sigma(\nu)}{\partial\nu} h(\nu)    \right)\nonumber \\
       &=& -k_BT\int_0^1 d\nu \frac{\partial S(\sigma(\nu))}{\partial\nu}  \nonumber\\
       &=& k_BT S(\sigma(0))\quad.
       \label{eq:Series_Limit1}
\end{eqnarray}
Using $h(0)=Q-q_0\mathds{1}$, it can be seen that $\Delta$
approaches to $0$ in this limit. This shows that $\Delta$ can be
chosen to be as small as possible. Moreover, for any given
parameter sequence $\{\nu_k\}$ it is possible to adjust the large
$k$ behavior of these sequences and make $Z_B$ finite. For
example, such an adjustment can be made only for indices $k$ that
exceed a threshold $k_{\mathrm{min}}$. By selecting the threshold
$k_{\mathrm{min}}$ sufficiently large, it is possible to insure
that $\Delta$ does not change much. In summary, therefore,
$\Delta$ can be made as small as possible by keeping $Z_B$ finite.

Since $\Delta$ depends continuously on $\nu_k$, it is then
possible to adjust these parameter sequences such that $\Delta$
has the value $\Delta =k_BTJ$, which is a strictly positive
quantity by the initial supposition. Then, by
Eq.~(\ref{eq:DeltaEB_expr_Land}), the change in the energy of the
bath is $\Delta E_B=\tr\,\rho Q$ and hence $Q$ is the HTO
associated with this realization. This completes the proof of the
Theorem \ref{cor:Landauer}.

Note that if the finiteness restriction on the bath is relaxed,
then the strict inequalities of (ii) and (iii) of Theorem
\ref{cor:Landauer} will be a non-strict inequality. The
construction of the realization above can also be adapted to the
special case of $J(\beta Q)=0$. In this case, all subsystems of B
are in the same $\sigma_1$ state and $Z_B$ is necessarily
infinite. Unfortunately, the correct energy change of B cannot be
calculated directly in such a state. At first sight, it appears
that the isometry changes the state of only the subsystem X$_1$ of
B. From here, one may try to compute the change in energy of B as
$\Delta E_B=\tr(\rho-\sigma_1)H_{X_1}$, which can easily be seen
to be incorrect. There is a subtle energy contribution of the
shift of information in an infinite chain and that contribution
can only be computed correctly as in the proof above by using a
bath with finite $Z_B$.

\section{Sufficiency Proof of Theorem \ref{thm:complete}}
\label{sec:app_com}

This section contains the proof of the sufficiency of the
conditions in Theorem \ref{thm:complete}. A separate proof has to
be given for the two conditions. For the special case of (ii), a
bath with a finitely many levels is sufficient while for (i), B
must necessarily have infinite levels. The notation below follow
the convention used in the proof of Theorem \ref{cor:Landauer}.

\textit{Sufficiency of (ii)}: Let $Q$ be an operator that
satisfies the conditions in (ii). Since $\sigma_{\beta Q}$ is
isospectral with $\rho_0$, there is a unitary operator $W$ on A
such that
\begin{equation}
  \beta Q= -W^\dagger (\ln \rho_0 +S(\rho_0)\mathds{1}_A) W \quad.
\end{equation}
For constructing the desired realization, the bath B will be
chosen to be a system identical with A and its equilibrium state
will be chosen to be $\rho_0$. The realization is obtained by
first transforming the device by $W$ and then swapping the states
of A and B. In other words, the realization isometry is
$U_{AB}=\mathcal{S}(W \otimes\mathds{1}_B)$ where the swap
operator is defined to satisfy
$\mathcal{S}\ket{\alpha}_A\otimes\ket{\alpha^\prime}_B=
\ket{\alpha^\prime}_A\otimes\ket{\alpha}_B$ for all $\ket{\alpha}$
and $\ket{\alpha^\prime}$. If the initial state of the device is
$\rho$, then the initial and final states of AB are
\begin{eqnarray}
  \rho_{AB} &=& (\,\rho\,)_A\otimes \rho_0\quad,\\
  \rho_{AB}^\prime &=& (\rho_0)_A \otimes (W\rho W^\dagger)_B.
\end{eqnarray}
As a result, $\mathcal{E}(\rho)=\rho_0$ and the energy of the bath
has been changed by
\begin{eqnarray}
 \Delta E_B &=& \tr\, H_B(W\rho W^\dagger  - \rho_0) \\
        &=&-k_BT \tr\,\ln \rho_0 (W \rho W^\dagger  - \rho_0)\\
        &=& -k_BT \tr\,\rho(W^\dagger \ln\rho_0 W +S(\rho_0)\mathds{1}_B) \\
        &=& \tr\,\rho Q
\end{eqnarray}
and therefore the HTO associated with this realization is $Q$.
This completes the proof of sufficiency of the special case (ii).

\textit{Sufficiency of (i)}: Let $Q$ be a hermitian operator
satisfying only the strict inequality in the statement (i). The
construction of the realization is quite similar to the
construction used in the proof of Theorem \ref{cor:Landauer}, but
now the bath has more components.

Let $d$ be the dimension of the Hilbert space of the device A and
let $r$ be the matrix rank of $\rho_0$.\footnote{It should be
mentioned that the dimension $d$ of the Hilbert space of A can be
smaller that $r$. In such cases, the reader may imagine that the
initial state $\rho_A$ of the device is always chosen from a
$d$-dimensional subspace of the Hilbert space of A. With this
interpretation, the HTO is an operator defined only on this
subspace.}. Let X be a $d$-level system and Y be an $r$-level
system. The bath is taken to be composed of infinitely many copies
of X and Y. The constitution of B can be expressed as
$\mathrm{B}=(\mathrm{Y}_1\mathrm{Y}_2\mathrm{Y}_3\cdots)(\mathrm{X}_1\mathrm{X}_2\mathrm{X}_3\cdots)$.
Let the standard orthonormal bases of these two types of
subsystems be denoted by the common notation $\{\ket{i}\}$ where
$i=0,1,\ldots,d-1$ for X and $i=0,1,\ldots,r-1$ for Y subsystems.
The following shorthand will be used for the state vectors of the
bath
\begin{equation}
  \ket{j_1j_2\cdots;i_1i_2\cdots}_B
          =    \left(\ket{j_1}_{Y_1}\otimes\ket{j_2}_{Y_2}\otimes\cdots\right)
               \otimes  \left(\ket{i_1}_{X_1}\otimes\ket{i_2}_{X_2}\otimes\cdots\right)  ~~.
\end{equation}
The Hilbert space $\mathcal{H}_B$ of the bath is defined as the
linear span of these states where the associated sequences
$\{j_k\}$ and $\{i_k\}$ contain only finitely many nonzero
elements. These sequences can be considered as the digits of
base-$r$ and base-$d$ representation of two nonnegative integers
$m$ and $n$ respectively. Namely, $m=(\cdots j_3j_2j_1)_r$ and
$n=(\cdots i_3i_2i_1)_d$. Therefore, $\mathcal{H}_B$ has a
countable basis and it is a separable Hilbert space.

\begin{figure}
\includegraphics[scale=0.8]{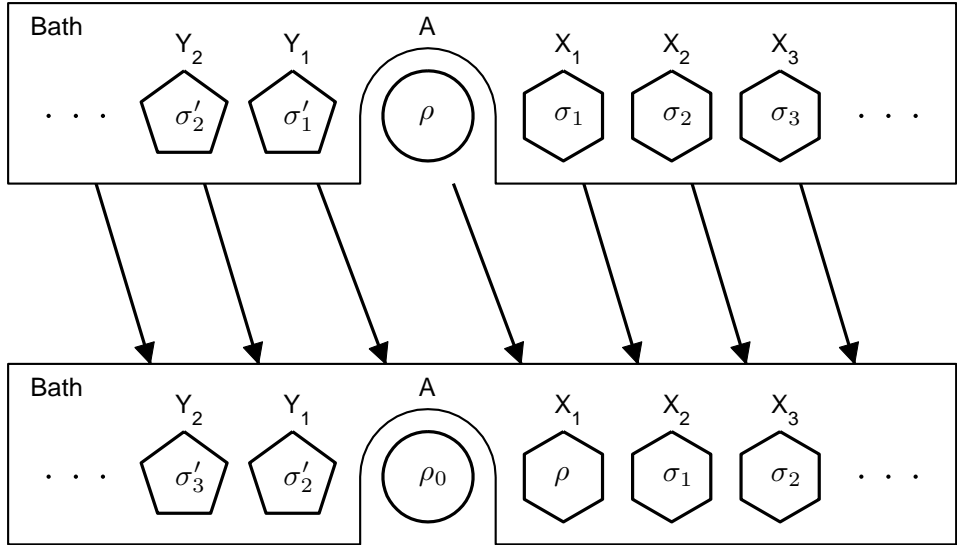}
\caption{The isometry used in the realization for the complete
erasure in the proof of the Theorem \ref{thm:complete}. The
initial state in the top row is changed to the final state in the
bottom row by copying the content of each subsystem to the next
one on the right. } \label{fig:complete}
\end{figure}

Let $\{\ket{i}\}$ and $\{\ket{\alpha_i}\}$ be the orthonormal sets
in the Hilbert space of the device A that appear in the spectral
decompositions of $Q$ and $\rho_0$ as in
Eq.~(\ref{eq:Q_spectral_decomp}) and
\begin{eqnarray}
 \rho_0 = \sum_{i=0}^{r-1} p_i\ket{\alpha_i}\bra{\alpha_i}    ~~ (p_0\geq p_1\geq\cdots\geq p_{r-1}>0)
\end{eqnarray}
where the eigenvalues are ordered in the indicated way. The
isometry $U_{AB}$ on the composite system is defined as follows
\begin{equation}
  U_{AB}\ket{\ell}_A \otimes \ket{j_1j_2j_3\cdots;i_1i_2i_3\cdots}_B
    = \ket{\alpha_{j_1}}_A \otimes \ket{j_2j_3j_4\cdots; \ell i_1i_2\cdots}_B~~,
\end{equation}
for any allowed values of input labels. Essentially, the isometry
$U_{AB}$ shifts the contents of the device and the subsystems in a
chain which is infinite in two directions. The movement of the
information among the subsystems is depicted in
Fig.~\ref{fig:complete}. Specifically,
\begin{itemize}
  \item[(i)] the content of Y$_k$ is copied to Y$_{k-1}$  for each $k\geq2$,
  \item[(ii)] the content of Y$_1$ is copied to A in such a way
  that the standard basis of Y$_1$ is carried onto the
  eigenvectors of $\rho_0$,
  \item[(iii)] the content of A is copied to X$_1$ in such a way
  that the eigenvectors of $Q$ is carried onto the standard basis
  of X$_1$,
  \item[(iv)] and the content of X$_k$ is copied onto X$_{k+1}$
  for all $k$.
\end{itemize}
Note that $U_{AB}$ maps an orthonormal basis of AB to an
orthonormal set and therefore it is an isometry.

For the Hamiltonian, the subsystems are taken to be independent
and the Hamiltonians of individual subsystems are described in
terms of a continuous parameter. For this purpose, consider
continuous functions $\epsilon_i(\nu)$ ($i=0,1,\cdots,d-1$) and
$\epsilon^\prime_j(\nu)$ ($j=0,1,\cdots,r-1$) of a real parameter
$\nu\in[0,1]$, to be used for energies of X and Y type subsystems
respectively. They are chosen to satisfy the following properties.
\begin{itemize}
\item[(i)] The zeroth level has zero energy,
$\epsilon_0(\nu)=\epsilon^\prime_0(\nu)=0$, for all $\nu$.
\item[(ii)] The energies of all excited levels diverge to
$+\infty$ as the corresponding parameter approaches to $1$, i.e.,
\begin{equation}
 \lim_{\nu\rightarrow1} \epsilon_i(\nu) =  \lim_{\nu\rightarrow1} \epsilon^\prime_j(\nu) = \infty\quad(\textrm{for~any}~i,j\geq1).
\end{equation}
\item[(iii)] The energy spectrum $\epsilon_i(\nu=0)$ is identical
with the eigenvalues of $Q$ up to a constant shift as
$\epsilon_i(0)=q_i-q_0$ for all $i$.
\item[(iv)] The energy spectrum $\epsilon^\prime_j(\nu=0)$ is such
that the equilibrium density matrix is isospectral with $\rho_0$,
i.e., $\epsilon^\prime_j(0)=k_BT\ln(p_0/p_j)$.
\end{itemize}

Let $\{\nu_k\}$ and $\{\nu^\prime_k\}$ be two infinite, increasing
sequences starting from $0$ and converging to $1$. The energy
levels of X$_k$ and Y$_k$ will be taken as $\epsilon_i(\nu_k)$ and
$\epsilon^\prime_i(\nu^\prime_k)$ respectively. In other words, in
terms of
\begin{eqnarray}
  h(\nu) &=& \sum_{i=0}^{d-1} \epsilon_i(\nu)\ket{i}\bra{i}\quad,\\
  h^\prime(\nu) &=& \sum_{i=0}^{r-1} \epsilon^\prime_i(\nu)\ket{i}\bra{i}\quad,
\end{eqnarray}
the Hamiltonians of individual systems can be expressed as
$H_{X_k}=h(\nu_k)$ and $H_{Y_k}=h^\prime(\nu^\prime_k)$. The bath
Hamiltonian $H_B$ is the sum of the subsystem Hamiltonians and
therefore the energy eigenvalue of the bath state
$\ket{j_1j_2\cdots;i_1i_2\cdots}_B$ is
\begin{equation}
  E(\{j_k\},\{i_k\})= \sum_{k=1}^\infty \epsilon^\prime_{j_k}(\nu_k^\prime) + \epsilon_{i_k}(\nu_k)~~~.
\end{equation}

Let $\sigma(\nu)=\exp(-\beta h(\nu))/\zeta(\nu)$ and
$\sigma^\prime(\nu)=\exp(-\beta h^\prime(\nu))/\zeta^\prime(\nu)$
be parameter-dependent density matrices and $\zeta(\nu)$ and
$\zeta^\prime(\nu)$ be the corresponding partition functions. The
shorthand notations $\sigma_k=\sigma(\nu_k)$ and
$\sigma^\prime_k=\sigma^\prime(\nu^\prime_k)$ will be used for the
states of X$_k$ and Y$_k$ respectively. The partition function of
the bath is then given by
\begin{equation}
  Z_B = \prod_{k=1}^\infty \zeta(\nu_k)\zeta^\prime(\nu^\prime_k)~~.
\label{eq:ZB}
\end{equation}
Depending only on the large $k$ behavior of the sequences
$\{\nu_k\}$ and $\{\nu^\prime_k\}$, this product may converge or
diverge.

In the following expressions, the subscripts are used for
indicating the subsystem a given density matrix specifies to.
Suppose that the initial state of the device A is $\rho$. The
initial and final density matrices of the combined system are
given by the following self-explanatory expressions.
\begin{eqnarray}
  \rho_{AB} &=& \cdots (\theta_2)_{Y_2} \otimes  (\theta_1)_{Y_1}
              \otimes (\rho)_A \otimes (\sigma_1)_{X_1} \otimes  (\sigma_2)_{X_2}  \cdots  \nonumber \\
  \rho_{AB}^\prime &=& \cdots (\theta_3)_{Y_2} \otimes  (\theta_2)_{Y_1}
              \otimes (\rho_0)_A \otimes (\rho)_{X_1} \otimes  (\sigma_1)_{X_2}  \cdots   \nonumber
\end{eqnarray}
It is obvious that the reduced quantum operation on the device is
$\mathcal{E}(\rho)=\rho_0$, i.e., the desired complete erasure.
The change in the average energy of the bath is given by
\begin{equation}
  \Delta E_B  = \tr (\rho Q) - k_BT(J+S(\rho_0)) + \Delta  \label{eq:DeltaEB_expr}
\end{equation}
where $J=J(\beta Q)$ and
\begin{eqnarray}
   \Delta &=& k_BT(J+S(\rho_0)) - \tr\,\sigma_1Q
                       + \sum_{k=2}^\infty \tr  (\sigma_{k-1}-\sigma_k) H_{X_k}
                       \nonumber\\
        & & + \sum_{k=1}^\infty \tr (\sigma^\prime_{k+1}-\sigma^\prime_k) H_{Y_k}~~,
   \label{eq:Delta_vs_DensMat}
\end{eqnarray}
which is a constant independent of the device state $\rho$.

When Eq.~(\ref{eq:magic}) is used, $\Delta$ can be expressed as
\begin{equation}
  \Delta = k_BT\sum_{k=1}^\infty
     \Big(  S(\sigma_k\vert\vert\sigma_{k+1})    +   S(\sigma^\prime_{k+1}\vert\vert\sigma^\prime_k) \Big)~~,
  \label{eq:Delta_vs_RelEnt}
\end{equation}
which shows that $\Delta$ is a strictly positive number. It
will be argued below that the sequences of subsystem parameters
$\{\nu_k\}$ and $\{\nu^\prime_k\}$ can be selected in such a way
that $\Delta$ can be adjusted to be equal to any positive value
and this can be achieved with a finite $Z_B$. As in the proof of
Theorem \ref{cor:Landauer}, this will be done by showing that
$\Delta$ can be made arbitrarily large and arbitrarily small.

(1) First, if the parameter $\nu_2$ of X$_2$ tends to $1$, the
final energy of X$_2$ approaches to infinity. In this limit
$\Delta$ diverges to $\infty$. (ii) In the limit where all
increments $\nu_{k+1}-\nu_k$ and $\nu^\prime_{k+1}-\nu^\prime_k$
tend to 0, the series in Eq.~(\ref{eq:Delta_vs_DensMat}) approach
to integrals which can readily be evaluated as in Eq.~(\ref{eq:Series_Limit1}) and
\begin{equation}
   \sum_{k=1}^\infty \tr (\sigma^\prime_{k+1}-\sigma^\prime_k) H_{Y_k}
       \longrightarrow -k_BT S(\sigma^\prime(0))~~.
\end{equation}
Therefore, in this limit, $\Delta$ approaches to $0$. This shows
that $\Delta$ can be chosen as small as possible. It can also be
shown that the same can be done with a finite $Z_B$ value.

Since $\Delta$ depends continuously on $\nu_k$ and $\nu^\prime_k$,
it is then possible to adjust these parameter sequences such that
$\Delta$ has the value
\begin{equation}
  \Delta = k_BT (J+S(\rho_0))\quad,
\end{equation}
which is a strictly positive quantity by our initial assumption.
In that case, $\Delta E_B=\tr\,\rho Q$ and hence $Q$ is the HTO
associated with this realization. This completes the proof of the
Theorem \ref{thm:complete}. $\Box$

\end{document}